\documentclass[aps,prd,nofootinbib,reprint,twocolumn,superscriptaddress,eqsecnum]{revtex4-2}

\usepackage[utf8]{inputenc}
\usepackage{graphicx}
\usepackage[caption=false]{subfig}
\usepackage{amsthm, amsfonts, amsmath, mathrsfs}
\usepackage{bbm}
\usepackage{slashed}
\usepackage{physics}
\usepackage{xspace}
\usepackage{orcidlink}
\usepackage{cancel}
\usepackage{comment}
\usepackage{glossaries}
\setacronymstyle{long-short}
\glsdisablehyper

\usepackage{hyperref}
\hypersetup{colorlinks=true,
    linkcolor=magenta,
    citecolor=blue,
    urlcolor=cyan,
    pdfpagemode=FullScreen,}

\newacronym{pionless}{EFT$_\slashed \pi$}{pionless effective field theory}
\newacronym{eft}{EFT}{effective field theory}
\newacronym{qed}{QED}{quantum electrodynamics}
\newacronym{qcd}{QCD}{quantum chromodynamics}
\newacronym{nrqed}{NRQED}{nonrelativistic QED}
\newacronym{vnrqed}{vNRQED}{velocity NRQED}
\newacronym{nrqcd}{NRQCD}{nonrelativistic QCD}
\newacronym{ChiPT}{\ensuremath{\chi}PT}{chiral perturbation theory}
\newacronym{ChiEFT}{\ensuremath{\chi}EFT}{chiral effective field theory}
\newacronym{BSM}{BSM}{Beyond the Standard Model}
\newacronym{NN}{\ensuremath{N\!N}\xspace}{nucleon-nucleon}
\newacronym{vRG}{vRG}{velocity renormalization group}
\newacronym{lo}{LO}{leading order}
\newacronym{nlo}{NLO}{next-to-leading order}
\newacronym{n2lo}{\ensuremath{\text{N}^2\text{LO}}}{next-to-next-to-leading order}
\newacronym{n3lo}{\ensuremath{\text{N}^3\text{LO}}}{next-to-next-to-next-to-leading order}
\newacronym{n4lo}{\ensuremath{\text{N}^4\text{LO}}}{next-to-next-to-next-to-next-to-leading order}
\newacronym{dimreg}{DimReg}{dimensional regularization}
\newacronym{lec}{LEC}{low energy coefficient}
\newacronym{av18}{AV18}{Argonne \ensuremath{v18}}
\newacronym{nEDM}{nEDM}{neutron electric dipole moment}

\newcommand{\sufour}{\ensuremath{{\rm SU}(4)}\xspace}

\newcommand{\SUtwo}{\ensuremath{{\rm SU}(2)}\xspace}

\newcommand{\Nc}{\ensuremath{N_c}\xspace}

\newcommand{\calH}{\ensuremath{\mathcal{H}}}

\newcommand{\calD}{\ensuremath{\mathcal{D}}}
\newcommand{\calL}{\ensuremath{\mathcal{L}}}
\newcommand{\calM}{\ensuremath{\mathcal{M}}}

\newcommand{\calP}{\ensuremath{\mathcal{P}}}
\newcommand{\1}{\mathbbm{1}}

\newcommand{\NN}{\ensuremath{N\!N}\xspace}

\begin{document}

\title{Strong CP violation and large-\Nc spin-flavor symmetry}

\author{Thomas R.~Richardson\,\orcidlink{0000-0001-6314-7518}}
\email{thomas.richardson@berkeley.edu}
\affiliation{Department of Physics, University of California, Berkeley, CA 94720, USA}
\affiliation{Nuclear Science Division, Lawrence Berkeley National Laboratory, Berkeley, CA 94720, USA}

\begin{abstract}
    We revisit the contribution of the QCD $\bar \theta$ term to the CP violating pion-nucleon couplings and the nucleon electric dipole moment in a combined large-$N_c$ and chiral perturbation theory framework.
    In particular, we approach this issue through the emergent spin-flavor symmetry of the baryon sector at large but finite $N_c$.
    We obtain good agreement with previous analyses for the pion-nucleon couplings and show that the large-$N_c$ framework indicates that tree-level contributions to the electric dipole moment possibly play a dominant role.
    The spin-flavor symmetry also enables us to provide novel constraints on CP violating pion-$\Delta$ couplings as well as the $\Delta$ electric dipole moment and $\Delta N$ transition moment.
\end{abstract}

\maketitle

\section{Introduction}

A key ingredient for baryogenesis is the violation of CP symmetry ($\cancel{\rm CP}$), where C is charge conjugation and P is parity \cite{sakharovViolationOfCPinVarianceCasymmetry1991}.
In the Standard Model, the phase of the Cabibbo-Kobayashi-Maskawa matrix provides a source of CP violation that suitably explains off-diagonal flavor transitions, but its impact is too small to generate flavor diagnoal CP violation necessary for baryogenesis \cite{gavelaSTANDARDMODELCPVIOLATION1994, gavelaStandardModelCPviolation1994a, huetElectroweakBaryogenesisStandard1995, konstandinAxialCurrentsCKM2004} (for a review, see Ref.~\cite{morrisseyElectroweakBaryogenesis2012}).

Another possible source of CP violation in the Standard Model is the \gls{qcd} $\bar \theta$ term that arises from instantons in the \gls{qcd} vacuum \cite{belavinPseudoparticleSolutionsYangMills1975, thooftSymmetryBreakingBellJackiw1976, jackiwVacuumPeriodicityYangMills1976, callanStructureGaugeTheory1976}.
It has long been known that $\bar \theta$ induces a \gls{nEDM}, which was first estimated in Ref.~\cite{baluni$mathrmCP$nonconservingEffectsQuantum1979} and assessed through current algebra techniques in Ref.~\cite{crewtherChiralEstimateElectric1979}.
While $\bar \theta$ could in principle be an $O(1)$ number, the lack of an observation of an \gls{nEDM} \cite{abelMeasurementPermanentElectric2020} implies it is close to zero.
This apparent smallness of $\bar \theta$ is referred to as the strong CP problem.

Translating the experimental limit on the \gls{nEDM} into a bound on $\bar \theta$ of course requires a careful theoretical analysis.
Over the years, there have been many calculations of the contribution to the \gls{nEDM} from the \gls{qcd} $\bar \theta$ term in ${\rm SU}(2)_L \times {\rm SU}(2)_R$ \gls{ChiPT} \cite{hockingsElectricDipoleForm2005, mereghettiEffectiveChiralLagrangian2010,mereghettiElectricDipoleForm2011}.
Additionally, there are several results based on \gls{qcd} sum rules \cite{pospelovThetaInducedElectricDipole1999, pospelovThetaVacuaQCD2000, pospelovElectricDipoleMoments2005, emaChiralPropertiesNucleon2024, hisanoReevaluationNeutronElectric2012}, the Skyrme model \cite{schnitzerSoftpionSkyrmionLagrangian1984, riggsCPviolatingYukawaCouplings1993, dixonElectricDipoleMoment1991}, the bag model \cite{morganNeutronElectricDipole1986, musakhanovElectricDipoleMoment1984}, and holographic \gls{qcd} \cite{hongElectricDipoleMoment2007, bartoliniNeutronElectricDipole2017, bartoliniThetaDependenceHolographic2017}.
More recently, the lattice \gls{qcd} community has presented first results for the \gls{nEDM} from $\bar \theta$ \cite{dragosConfirmingExistenceStrong2021, bhattacharyaContributionQCD$Th$term2021, liangNucleonElectricDipole2023, alexandrouNeutronElectricDipole2021} (see Ref.~\cite{liuLatticeQCDNeutron2025} for a recent review).
The chiral and continuum extrapolations/interpolations in these lattice calculations, are generally based on the ${\rm SU}(2)_L \times {\rm SU}(2)_R$ \gls{ChiPT} results.
Despite this tremendous progress, we should recall that the presence of the $\bar \theta$ term is intimately connected with the ${\rm U(1)}_A$ anomaly \cite{thooftSymmetryBreakingBellJackiw1976, adlerAbsenceHigherOrderCorrections1969, adlerAxialVectorVertexSpinor1969, bellPCACPuzzleP0-gg1969}.
Therefore, it seems reasonable to expect that a formulation of \gls{ChiPT} that takes the anomaly into account explicitly could have some advantages.

There are also aspects of the large-\Nc limit of \gls{qcd} \cite{thooftPlanarDiagramTheory1974}, where \Nc is the number of colors, that place additional constraints on the structure of the effective theory with the axial anomaly incorporated explicitly.
First, it has been shown that the pattern of chiral symmetry breaking is lifted to ${\rm U}(N_f)_L \times {\rm U}(N_f)_R \to {\rm U}(N_f)_V$, where $N_f$ is the number of light quark flavors, in the large-\Nc limit because the axial anomaly is 1/\Nc suppressed \cite{colemanChiralSymmetryBreakdownLarge1980}.
Thus, there is an additional Goldstone mode in the chiral and large-\Nc limits corresponding to the spontaneously broken ${\rm U}(1)_A$.
At finite \Nc and at lowest order in the chiral limit, the axial anomaly generates a mass for the flavor-singlet meson that is tied to the topological susceptibility of pure Yang-Mills, which is captured by the famous Witten-Veneziano formula \cite{venezianoU1Instantons1979, wittenCurrentAlgebraTheorems1979}.
These constraints were implemented in the mesonic chiral Lagrangian for the meson sector several decades ago \cite{divecchiaChiralDynamicsLarge1980,divecchiaEffectiveLagrangianNo1979,wittenCurrentAlgebraTheorems1979,wittenLargeChiralDynamics1980,venezianoU1Instantons1979, kaiserLargeChiralPerturbation2000, rosenzweigEffectiveLagrangianQuantum1980, Ohta:1981ai, Kawarabayashi:1980dp, Kawarabayashi:1980uh}.

There have also been several studies that extended the three-flavor baryon chiral Lagrangian to include the effects of the anomaly with certain appeals to the large-\Nc limit \cite{pichStrongCPviolationEffective1991, borasoyElectricDipoleMoment2000, ottnadNewInsightsNeutron2010, aokiStrongCPViolation1992}.
While Refs.~\cite{pichStrongCPviolationEffective1991, borasoyElectricDipoleMoment2000, ottnadNewInsightsNeutron2010, aokiStrongCPViolation1992} correctly take into account the constraints of the large-\Nc limit in the meson sector, there are additional constraints in the baryon sector that emerge in the large-\Nc limit and were not incorporated into the effective theory.
In particular, there is an emergent ${\rm SU}(2 N_f)$ spin-flavor symmetry that relates the nucleons and the $\Delta$ resonances in the two-flavor case and octet and decuplet baryons in the three-flavor case \cite{dashenBaryonPionCouplings1993, dashenExpansionBaryons1994, dashenSpinFlavorStructure1995, gervaisLargeBaryonicSoliton1984, gervaisLargeQCDBaryon1984, caroneSpinIndependenceLarge1994, lutyBaryonsQuarksExpansion1994}.
This symmetry provides one explanation for the important role of decuplet baryons in loop corrections to octet baryon properties in \gls{ChiPT} \cite{jenkinsChiralCorrectionsBaryon1991, flores-mendietaStructureLargeCancellations2000}.
An economical Lagrangian that incorporates the effects of spin-flavor symmetry and chiral symmetry simultaneously was developed in Ref.~\cite{jenkinsChiralLagrangianBaryons1996}, which we will adopt in this work.

In light of the various appeals to large-\Nc in \gls{nEDM} estimates and the option to treat the axial anomaly explicitly, it is appropriate to investigate the consistency of the different approaches.
We will show:
    \begin{itemize}
        \item the combination of chiral symmetry with the $1/\Nc$ expansion \cite{divecchiaChiralDynamicsLarge1980,divecchiaEffectiveLagrangianNo1979,wittenLargeChiralDynamics1980,pichStrongCPviolationEffective1991,jenkinsBaryonChiralPerturbation1991, Ohta:1981ai, Kawarabayashi:1980dp, Kawarabayashi:1980uh} and the Witten-Veneziano formula \cite{venezianoU1Instantons1979, wittenCurrentAlgebraTheorems1979} gives $\cancel{\rm CP}$ pion-nucleon couplings consistent with previous estimates based on ${\rm SU}(2)_L \times {\rm SU}(2)_R$ \gls{ChiPT},

        \item the \sufour spin-flavor symmetry predicts the ratio of $\cancel{\rm CP}$ pion-$\Delta$ couplings to the $\pi \NN$ couplings up to $O(1/\Nc^2)$ corrections,

        \item the spin-flavor symmetry renders the \gls{ChiPT} loop expansion for the \gls{nEDM} as a $1/\Nc$ expansion such that the finite part of the tree-level contribution is possibly dominant compared to the chiral loop,

        \item the spin-flavor symmetry allows us to derive isospin relations for the neutron, proton, and $\Delta$ EDMs as well as $\Delta N$ transition EDMs that are identical to magnetic moment relations.
    \end{itemize}
Throughout this work, an important theme is that consistent large-\Nc estimates require the explicit inclusion of the ${U}(1)_A$ anomaly before analyzing the large-\Nc scaling of certain couplings in addition to the constraints of the spin-flavor symmetry.
Additional formulae for the matrix elements of the spin-flavor generators and loop integrals can be found in the Appendix.

\section{Strong CP violation in the $1/\Nc$ chiral Lagrangian}
    \label{sec:lagrangian}

\subsection{Meson Lagrangian}
    \label{subsec:meson_lagrangian}

In ${\rm SU}(\Nc)$ \gls{qcd} with $N_f$ light flavors, the axial anomaly is $1/\Nc$ suppressed such that the pattern of spontaneous chiral symmetry breaking at large-\Nc is ${\rm U}(N_f)_L \times {\rm U}(N_f)_R \to {\rm U}(N_f)_V$ \cite{colemanChiralSymmetryBreakdownLarge1980}.
In this work we consider $N_f = 2$.
The ${\rm U}(1)_A$ anomaly can be taken into account explicitly by including a term in the Lagrangian that breaks ${\rm U}(1)_A$ explicitly but preserves $\SUtwo_L \times \SUtwo_R \times {\rm U}(1)_V$.
The lowest order effective Lagrangian in the meson sector has been derived several times \cite{divecchiaChiralDynamicsLarge1980, divecchiaEffectiveLagrangianNo1979, rosenzweigEffectiveLagrangianQuantum1980, wittenCurrentAlgebraTheorems1979, wittenLargeChiralDynamics1980, pichStrongCPviolationEffective1991, Ohta:1981ai, Kawarabayashi:1980dp, Kawarabayashi:1980uh},
    \begin{align}
        \calL_{\pi} & = \frac{F_0^2}{4} \Tr \left( D_\mu \bar U (D^\mu \bar U)^\dagger \right) + \frac{F_0^2}{4} \Tr \left( \bar U^\dagger \tilde \chi + \tilde \chi^\dagger \bar U \right) \nonumber \\
        & - \frac{F_0^2}{4} \frac{a}{\Nc} \left[ \frac{i}{2} \left( \log \det \bar U - \log \det \bar U^\dagger \right) \right]^2 \, ,
        \label{eq:meson_lagrangian_1}
    \end{align}
where $F_0 \sim \sqrt{\Nc}$ is the pion decay constant in the chiral limit.
In ordinary \gls{ChiPT}, the ground state is parameterized by the identity matrix, i.e., $\langle \bar U \rangle = \1$.
However, this is no longer true in an arbitrary $\theta$ vacuum.
Instead, we have $\bar U = \langle \bar U \rangle U$ where the ground state is given by
    \begin{equation}
        \langle \bar U \rangle = \begin{pmatrix}
            e^{i \varphi_u} & 0 \\
            0 & e^{i \varphi_d}
        \end{pmatrix} \, ,
    \end{equation}
and the pseudo-Nambu-Goldstone modes are contained in the matrix $U$,
    \begin{equation}
        U = e^{\frac{i}{F_0} \left(\phi_0 \1 + \phi_a \tau^a \right)} \, ,
    \end{equation}
where $\tau^a$ are the Pauli matrices in isospin space with $a = 1, 2, 3$.

The contributions from the quark masses are contained in $\tilde \chi = 2 B_0 \calM$, where $\calM$ is a complex valued mass matrix and initially contains all of the $\bar \theta$ dependence.
Through a suitable choice of transformations, $\calM$ can be brought into the form
    \begin{align}
        \calM = e^{-i \bar \theta/2} M \, ,
    \end{align}
where $M = {\rm diag}(m_u, m_d)$ is real and diagonal.
This dependence can be removed from the mass term with the ${\rm U}(1)_A$ transformation,
    \begin{align}
        \bar U \mapsto e^{-i \bar \theta/2} \bar U \, .
    \end{align}
This transformation only modifies the logarithmic terms in Eq.~\eqref{eq:meson_lagrangian_1}.
Additionally, we can use the decomposition $\bar U = \langle \bar U \rangle U$ and introduce $\widetilde M(\theta) = {\rm diag} \left( m_u \cos \varphi_u, \ m_d \cos \varphi_d \right)$, which leaves us with the final form for the mesonic Lagrangian,
    \begin{equation}
        \begin{split}
            \calL_\pi & = \frac{F_0^2}{4} \Tr \left( D_\mu U (D^\mu U)^\dagger \right) + \frac{1}{2} B_0 F_0^2 \Tr \left( \widetilde M U + U^\dagger \widetilde M \right) \\
            & + \frac{F_0^2}{4} \frac{i a \tilde \theta}{\Nc} \Tr \left[ \left(U - U^\dagger\right)  - \log U + \log U^\dagger \right] \\
            & - \frac{F_0^2}{4} \frac{a}{\Nc} \left[\frac{i}{2} \left( \log \det U - \log \det U^\dagger \right) \right]^2 \, ,
        \end{split}
        \label{eq:meson_lagrangian}
    \end{equation}
where we have omitted an irrelevant constant proportional to $\tilde \theta^2$.
This Lagrangian yields the correct vacuum alignment to eliminate all tadpoles at lowest order in the chiral expansion \cite{crewtherChiralEstimateElectric1979}.
The term on the final line of Eq.~\eqref{eq:meson_lagrangian} sets the mass of $\phi_0$ in the chiral limit,
    \begin{align}
        m_0^2 = \frac{2 a}{\Nc} \, . \label{eq:isoscalar_meson_mass}
    \end{align}
In the three-flavor case, this additional meson can be interpreted as the $\eta'$; in the two-flavor case there is not a clear interpretation of the field in terms of a physical particle.
Rather, it can be considered some admixture of the $\eta$ and the $\eta'$.
Regardless of this interpretation, we consider $a \sim O(1)$ and remark that this term gives $\phi_0$ a mass in the chiral limit while it remains massless in the $\Nc \to \infty$ limit.

In the chiral limit, the mass of the isoscalar meson is also related to the topological susceptibility in Yang-Mills according to the Witten-Veneziano formula \cite{venezianoU1Instantons1979, wittenCurrentAlgebraTheorems1979}
    \begin{align}
        m_0^2 = \frac{2 N_f}{F_0^2} \chi_{\rm{YM}} \, .
    \end{align}
Combining this with Eq.~\eqref{eq:isoscalar_meson_mass} we fix $a/\Nc$,
    \begin{align}
        \frac{a}{\Nc} = \frac{N_f}{F_0^2} \chi_{\rm{YM}} \, . \label{eq:aNc_YM}
    \end{align}
We take an average of several lattice calculations \cite{cichyNonperturbativeTestWittenVeneziano2015, durrPrecisionStudySU32007, deldebbioTopologicalSusceptibilitySU32005, luscherUniversalityTopologicalSusceptibility2010, ceNonGaussianitiesTopologicalCharge2015, chowdhuryTopologicalSusceptibilityLattice2014, athenodorouGlueballSpectrumSU32020, bonati$th$Dependence$SU3$2016} to obtain
    \begin{equation}
        \chi^{1/4}_{\rm YM} = 188.8(1.8) \ {\rm MeV} \, .
    \end{equation}
Also, we use the FLAG estimate of the pion decay constant in the chiral limit $F_0 = 86.2(5) \ {\rm MeV}$ and at the physical pion mass $F_\pi = 92.3$ MeV \cite{aokiFLAGReview20212022, aokiFLAGReview20242025}, which leads to $m_0 = 827(17) \ {\rm MeV}$ when $N_f = 2$ or equivalently $a/\Nc = ( 585(12) \ {\rm MeV})^2$.
Furthermore, it is worth noting that an explicit comparison of the $\eta'$ mass and the topological susceptibility of pure Yang-Mills appears to validate the Witten-Veneziano formula \cite{cichyNonperturbativeTestWittenVeneziano2015}.

Lastly, the angles $\varphi_u$ and $\varphi_d$ are determined by minimizing the effective potential, which gives the condition
    \begin{align}
        2 B_0 m_i \sin \varphi_i = \frac{a}{\Nc} \tilde  \theta \, , \label{eq:ground_state}
    \end{align}
with $\tilde \theta = \bar \theta - \sum_j \varphi_j$.
Away from the chiral limit in the realistic situation where $m_u , \, m_d \ll a/\Nc$, the solution is \cite{wittenLargeChiralDynamics1980, divecchiaChiralDynamicsLarge1980}
    \begin{equation}
    \begin{split}
        \sin \varphi_u = \frac{m_d \sin \bar \theta}{\sqrt{m_u^2 + m_d^2 + 2 m_u m_d \cos \bar \theta}} \approx \frac{m_d \bar \theta}{m_u + m_d}  \, , \\
        \sin \varphi_d = \frac{m_u \sin \bar \theta}{\sqrt{m_u^2 + m_d^2 + 2 m_u m_d \cos \bar \theta}} \approx \frac{m_u \bar \theta}{m_u + m_d} \, .
    \end{split}
    \label{eq:ground_state_solution}
    \end{equation}
Above, we have made use of the phenomenological fact that $\bar \theta \ll 1$.
Combining Eqs.~\eqref{eq:ground_state} and \eqref{eq:ground_state_solution} gives the relation
    \begin{equation}
        2 B_0 m_* \bar \theta = \frac{a}{\Nc} \tilde \theta \, , \label{eq:aOverNc_pion_mass1} 
    \end{equation}
where  
    \begin{align}
        m_* & = \frac{m_u m_d}{m_u + m_d} = \frac{1}{2} \bar m \left( 1 - \varepsilon^2 \right) \, , \label{eq:mstar} \\
        \bar m & = \frac{1}{2} \left( m_u + m_d \right) \, , \label{eq:quark_mass_avg} \\
        \varepsilon & = \frac{m_d - m_u}{m_u + m_d} \, . \label{eq:quark_mass_iso_break}
    \end{align}
Inserting Eq.~\eqref{eq:mstar} into Eq.~\eqref{eq:aOverNc_pion_mass1} and neglecting $\varepsilon$ shows 
    \begin{equation}
        \frac{a}{\Nc} \tilde \theta = B_0 \bar m \bar \theta \approx \frac{1}{2} m_\pi^2 \bar \theta \, , \label{eq:aOverNc_pion_mass2}
    \end{equation} 
which provides a motivation for the linked expansion in powers of $1/\Nc$ and $m_\pi^2$ advocated in Ref.~\cite{kaiserLargeChiralPerturbation2000}.
Using the estimate for the topological susceptibility leads to $\tilde \theta \approx \frac{m_\pi^2 / 2}{a/\Nc} \bar \theta \sim 0.026 \, \bar \theta$, which is consistent with the $N_f = 3$ estimate in Refs.~\cite{borasoyElectricDipoleMoment2000,ottnadNewInsightsNeutron2010}.

In this work, we will leave factors $\frac{a}{\Nc} \tilde \theta$ explicit until providing numerical estimates for various quantities. 
This is due to the fact that the substitution in Eq.~\eqref{eq:aOverNc_pion_mass2} can obscure the large-\Nc scaling of different couplings or observables since it implicitly sets the ordering of the chiral and large-\Nc limits, i.e., there is not a sense in which the large-\Nc limit can be applied after using Eq.~\eqref{eq:aOverNc_pion_mass2} since it assumes the pions are much lighter than the isospin-singlet meson.
Another way to view this is that the left-hand-side of Eq.~\eqref{eq:aOverNc_pion_mass2} is $O(1/\Nc)$ while the right-hand-side is $O(1)$, which will obscure any large-\Nc estimates.
This is similar to a result concerning the \gls{nlo} operators in the large-\Nc $N_f = 3$ chiral Lagrangian \cite{perisLargeNcBehaviourL71995}.

\subsection{Baryon Lagrangian}
    \label{subsec:baryon_lagrangian}

In the $\Nc \to \infty$ limit, a contracted ${\rm SU}(2 N_f)$ symmetry emerges in the baryon sector \cite{dashenBaryonPionCouplings1993, dashenExpansionBaryons1994, dashenSpinFlavorStructure1995, gervaisLargeBaryonicSoliton1984,gervaisLargeQCDBaryon1984, caroneSpinIndependenceLarge1994, lutyBaryonsQuarksExpansion1994}.
At large but finite $N_c$, the baryon matrix elements of any QCD operator containing $m$ quark bilinears can be expanded in terms of the SU(4) generators as \cite{dashenSpinFlavorStructure1995}
    \begin{equation}
            \label{eq:spin-flavor-expansion}
        \mathcal O^{(m)}_{QCD} = N_c^m  \sum_{n,s,t} c_n  \left( \frac{\hat J^i}{N_c} \right)^{s} \left( \frac{\hat I^a}{N_c} \right)^{t} \left( \frac{\hat G^{jb}}{N_c} \right)^{n - s - t} \, ,
    \end{equation}
where $c_n$ is an $O(1)$ coefficient that depends on nonperturbative \gls{qcd} dynamics, and the generators are
    \begin{align}
        \hat J^i & = q^\dagger \frac{\sigma^i}{2} q \, , \\
        \hat I^a & = q^\dagger \frac{\tau^a}{2} q \, , \\
        \hat G^{ia} & = q^\dagger \frac{\sigma^i \tau^a}{4} q \, .
    \end{align}
It should be understood that we are considering the baryon matrix elements $\langle B'| \mathcal O^{(m)}_{QCD} |B \rangle$ and $| B \rangle$ is a baryon state composed of \Nc totally symmetric indices in the fundamental representation of \sufour.
The matrices $\sigma^i$ ($\tau^a$) are the usual Pauli matrices in spin (isospin) space.
The matrix elements of the generators for physical baryons scale as
    \begin{equation}
        \begin{split}
        \langle B' | \hat J^i | B \rangle, \, \langle B' | \hat I^a | B \rangle & \sim O(N_c^0) \, , \\
        \langle B' | \hat G^{ia} | B \rangle & \sim O(N_c) \, .
        \end{split}
    \end{equation}
This so-called spin-flavor expansion was implemented in the chiral Lagrangian in Ref.~\cite{jenkinsChiralLagrangianBaryons1996}.
The lowest order Lagrangian is 
    \begin{equation}
        \begin{split}
            \calL_B & = i D_0 + \frac{1}{2} \Tr \left( u_i \tau^a \right) A^{i a} + \cdots
        \end{split} \, ,
    \end{equation}
where each term is understood to be bilinear in a ${\rm SU}(2 N_f)$ valued baryon field.
The vielbein is given by
    \begin{equation}
        u_i = i \left[ u^\dagger \partial_i u - u \partial_i u^\dagger \right] \, ,
    \end{equation}
where $u^2 = U$, and the axial current is given by
    \begin{align}
        A^{ia}_{B' B} & = \bra{B'} \bar q \gamma^i \gamma^5 \frac{\tau^a}{2} \ket{B} \, \\
        & = \bra{B'} a^{(v)}_1 G^{ia} + \frac{a^{(v)}_2}{\Nc} J^i I^a + \frac{a^{(v)}_3}{\Nc^2} \left\{ J^2, G^{ia} \right\} \ket{B} \, . \label{eq:spin-flavor-axial}
    \end{align}
The contributions from the quark mass matrix are \cite{pichStrongCPviolationEffective1991, jenkinsChiralLagrangianBaryons1996}
    \begin{align}
        \calL & = \Tr(\chi_+) \calH^0 +   \Tr(\chi_+ \tau^a) \calH^a \label{eq:baryon-mass-lagrangian}
    \end{align}
where 
    \begin{equation}
        \chi_+ = 2 B_0 \left( u^\dagger \widetilde M u^\dagger + u \widetilde M u \right) + \frac{i a \tilde \theta}{\Nc} \left( U - U^\dagger \right) \, . \label{eq:chi_+}
    \end{equation}
The baryon matrix elements are given by
    \begin{align}
        \calH^0_{B' B} & = \frac{1}{\Lambda} \bra{B'} \bar q q \ket{B} \nonumber \\
        & = \bra{B'} \Nc h^{(0)}_0 + \frac{1}{\Nc} h^{(0)}_2 J^2 \ket{B} \label{eq:spin-flavor-h0} \\ 
        \calH^a_{B' B} & = \frac{1}{\Lambda} \bra{B'} \bar q \frac{\tau^a}{2} q \ket{B} \nonumber \\
        & = h^{(v)}_1 I^a + \frac{1}{\Nc^2} h^{(v)}_3 J^2 I^a \, \label{eq:spin-flavor-ha}.
    \end{align}
In these two equations, $\Lambda$ is an arbitrary scale of dimension $1$, which is then absorbed into the coefficients $h_i^{(0,v)}$ of the spin-flavor expansion.
In principle, this scale could be taken to be the \gls{qcd} scale parameter $\Lambda_{\rm QCD}$ since this is a fixed quantity in the large-\Nc limit and the only dimensionful scale in \gls{qcd} apart from the quark masses.

Inserting the decomposition in Eq.~\eqref{eq:chi_+} leads to the $CP$ conserving mass terms
    \begin{align}
        \calL_M & = 2 B_0 \Tr(U^\dagger \widetilde M + \widetilde M U) \calH^0 \nonumber \\
        & + 2 B_0 \Tr \left( u^\dagger \widetilde M u^\dagger \tau^a + u \widetilde M u \tau^a \right) \calH^a \, .
        \label{eq:baryon_quark_mass}
    \end{align}
With this normalization the tree-level neutron-proton mass splitting $\delta m_N = m_n - m_p$ due to strong isospin breaking is
    \begin{equation}
        \begin{split}
            \delta m_N & = - 4 B_0 \frac{m_d^2 - m_u^2}{\sqrt{m_u^2 + m_d^2 + 2 m_u m_d \cos \theta}} \\
            & \times \left( h_1^{(v)} - \frac{3}{4 \Nc^2} h_3^{(v)} \right) \, .
        \end{split}
    \end{equation}
This reduces to
    \begin{align}
        \delta m_N & = -8 B_0 c_5 \bar m \varepsilon
    \end{align}
when $\theta = 0$ and identifying 
    \begin{equation}
        c_5 = h_1^{(v)} - \frac{3}{4 \Nc^2} h_3^{(v)} \label{eq:baryon_c5}
    \end{equation}
in agreement with Ref.~\cite{devriesBaryonMassSplittings2015, devriesLatticeQCDSpectroscopy2017}.
In order to draw clearer comparisons to Ref.~\cite{devriesBaryonMassSplittings2015}, we will adopt the same values for the average quark mass $\bar m$ and $\varepsilon$ from FLAG \cite{aokiReviewLatticeResults2014} and the weighted average for the nucleon mass splitting \cite{borsanyiInitioCalculationNeutronproton2015, beaneStrongIsospinViolationNeutronProton2007, blumElectromagneticMassSplittings2010, qcdsf-ukqcdcollaborationIsospinBreakingOctet2012, rm123collaborationLeadingIsospinBreaking2013, borsanyiIsospinSplittingsLight2013}
    \begin{align}
            \label{eq:mass_params}
        \bar m = 3.42(9) \, {\rm MeV} \, , \quad \varepsilon = 0.37(3) \, , \quad \delta m_N = 2.49(17) \, {\rm MeV} \, .
    \end{align}

The first term in the expansion of the isoscalar term in Eq.~\eqref{eq:baryon_quark_mass} can be identified with
    \begin{equation}
        c_1 = \Nc h^{(0)}_0 + \frac{3}{4 \Nc} h^{(0)}_2 \, .
    \end{equation}
It is well known that $c_1$ is related to the pion-nucleon sigma term.
We will use the estimate $c_1  = -1.0(3) \ {\rm GeV}^{-1}$ \cite{baruPrecisionCalculationThreshold2011}.

The decomposition in Eq.~\eqref{eq:chi_+} also induces the $\cancel{\rm CP}$ terms
    \begin{align}
        \calL_{\cancel{CP}} & = \frac{i a \tilde \theta}{\Nc} \Tr(U - U^\dagger) \calH^0 \nonumber \\
        & + \frac{i a \tilde \theta}{\Nc} \Tr(U \tau^a - U^\dagger \tau^a) \calH^a  \, ,
    \end{align}
which will give the $\cancel{\rm CP}$ baryon-pion couplings when expanded to $O(\phi)$,
    \begin{align}
      \calL_{\cancel{CP}}  & = - \frac{4 a \tilde \theta}{F_0 \Nc} \phi_0 \calH^0 - \frac{4 a \tilde \theta}{F_0 \Nc} \phi_a \calH^a \, .\label{eq:cp_violating_couplings}
    \end{align}
The first term gives gives an $O(\Nc^{-1/2})$ vertex while the second term gives a $O(\Nc^{-3/2})$ vertex.
Thus, the isovector coupling is $1/\Nc$ suppressed relative to the isoscalar coupling as expected where we are referring to the isospin of the baryon matrix element only.

This form of the Lagrangian can be matched onto that of Refs.~\cite{devriesBaryonMassSplittings2015, devriesLatticeQCDSpectroscopy2017},
    \begin{equation}
        \calL_{\cancel {\rm CP}} = - \frac{\bar g_0^{(\pi \NN)}}{2 F_0} \phi_a N^\dagger \tau^a N - \frac{ \bar g_1^{(\pi \NN)}}{2 F_0} \phi_3 N^\dagger N \, . \label{eq:matching_lagrangian}
    \end{equation}
When the spin-flavor generator of the isovector term is evaluated on nucleon states we find
    \begin{align}
        \frac{\bar g_0^{(\pi \NN)}}{2 F_0} & = \frac{2 a}{F_0 \Nc} \tilde \theta \left( h_1^{(v)} - \frac{3}{4 \Nc^2} h_3^{(v)} \right) \label{eq:g0} \, ,
    \end{align}
which is $O(\Nc^{-3/2})$.
Inserting Eq.~\eqref{eq:aOverNc_pion_mass2} and the isospin-breaking nucleon mass splitting Eq.~\eqref{eq:mass_params} gives
    \begin{align}
        \frac{\bar g_0^{(\pi \NN)}}{2 F_0} & = 0.7(1) \, \tilde \theta = 0.017(3) \, \bar \theta  \, . \label{eq:g0_estimate}
    \end{align}
This result is in agreement with the one based on \SUtwo \gls{ChiPT} \cite{mereghettiEffectiveChiralLagrangian2010, mereghettiElectricDipoleForm2011, hockingsElectricDipoleForm2005, devriesBaryonMassSplittings2015, devriesLatticeQCDSpectroscopy2017}.

The first term in Eq.~\eqref{eq:cp_violating_couplings} gives a $\cancel{\rm CP}$ coupling to the $\phi_0$ field, which mixes kinetically with $\pi^0$ when we are away from the isospin limit.
Integrating out the $\phi_0$ at tree-level gives
    \begin{equation}
        \phi_0 = \frac{\Nc}{a} B_0 \bar m \varepsilon \phi_3 \, .
    \end{equation}
Combining this with the spin-flavor matrix element evaluated on nucleon states and making use of Eq.~\eqref{eq:aNc_YM} leads to
    \begin{align}
        \frac{\bar g_1^{(\pi \NN)}}{2 F_0} & = \frac{F_0}{2 \chi_{\rm YM}} (2 B_0 \bar m)^2 \varepsilon (1 - \varepsilon^2) c_1 \bar \theta \, , \label{eq:g1}
    \end{align}
Finally, we have the numerical estimate
    \begin{align}
        \frac{\bar g_1^{(\pi \NN)}}{2 F_0} & =  -0.004(1) \, \bar \theta \, , \label{eq:g1_estimate}
    \end{align}
which is compatible with the value Refs.~\cite{bsaisouNuclearElectricDipole2015a, devriesIndirectSignsPecceiQuinn2019, devriesBaryonMassSplittings2015} before including the effects of the of the additional contribution from $\chi_-$ and the odd-parity nucleon resonance \cite{bsaisouElectricDipoleMoment2013}.
We also have $\bar g_1^{(\pi \NN)}/ \bar g_0^{(\pi \NN)} \sim - 0.2$, which is clearly not $O(\Nc)$.
This relative size can be attributed to the fact that we treated the $\phi_0$ as a heavy particle and integrated it out, while it should be considered as a light pseudo-Nambu-Goldstone boson in the large-\Nc limit.
Moreover, integrating out the $\phi_0$ produces an additional factor of $\varepsilon$.
Importantly, this illustrates a potential pitfall in constraining couplings from the spin-flavor expansion without consistently including the $\phi_0$ explicitly.
However, this result does extend the idea of Ref.~\cite{devriesLatticeQCDSpectroscopy2017} to constrain $\cancel{\rm CP}$ couplings from lattice \gls{qcd} spectroscopy; specifically, Eq.~\eqref{eq:g1} constrains $\bar g_1^{(\pi \NN)}$ in terms of the pion-nucleon sigma term \textit{and} $\chi_{\rm YM}$.

The large-\Nc scalings of $\bar g_0^{(\pi \NN)}$ and $\bar g_1^{(\pi \NN)}$, up to an overall factor of $F_0$, have also been determined by an analysis of the time-reversal-invariance violating \NN potential in chiral effective field theory \cite{samartTimereversalinvarianceviolatingNucleonnucleonPotential2016}.
There, the authors find $\bar g_0^{(\pi \NN)}/\bar g_1^{(\pi \NN)} \sim O(1/\Nc)$.
If we determined the large-\Nc scaling from Eq.~\eqref{eq:matching_lagrangian}, then the we would find agreement; indeed, this is the case for the chromoelectric and chromomagnetic dipole moments \cite{Bhattacharya:2025blb}.
However, we should recall that $\bar g_1^{(\pi \NN)}$ in Eq.~\eqref{eq:matching_lagrangian} was determined integrating out the $\phi_0$ field at tree-level, which introduces a factor of $\Nc \varepsilon/a$ in the definition of $\bar g_1^{(\pi \NN)}$ and upsets the manifest large-\Nc scaling.
Again, this highlights a possible pitfall in constraining couplings from spin-flavor symmetry without an explicit $\phi_0$ for the case of the $\bar \theta$ term while the analysis for higher-dimensional sources of $\cancel {\rm CP}$ does not appear to depend on this detail.

Reference~\cite{samartTimereversalinvarianceviolatingNucleonnucleonPotential2016} also includes an analysis of a potential containing more general meson-nucleon couplings.
In this case they find that $\bar g_0^{(\pi \NN)}$ is 1/\Nc suppressed relative to the $\cancel{\rm CP}$ coupling of the $\eta$ to the nucleon, i.e., $\bar g_0^{(\eta \NN)}$.
The structure of the couplings in Eq.~\eqref{eq:cp_violating_couplings} agrees with this conclusion.
Still, $\bar g_1^{(\pi \NN)}$ only arises from integrating out the $\phi_0$ as discussed above at the order we are considering, so the large-\Nc scaling relative to $\bar g_0^{(\pi \NN)}$ does not necessarily agree with the general potential in Ref.~\cite{samartTimereversalinvarianceviolatingNucleonnucleonPotential2016}.
Rather, it is likeley again to be dependent on the specific source of CP violation.

The spin-flavor symmetry also allows us to relate the $\cancel{\rm CP}$ $\pi \Delta \Delta$ couplings to the $\pi \NN$ couplings.
We will normalize the couplings $g_0^{(\pi \Delta \Delta)}$ such that they reproduce the expressions in Ref.~\cite{sengParityTimeReversalViolating2017} when expanded in terms of the physical fields.
At this order, there are no $\cancel{\rm CP}$ $\pi N \Delta$ couplings as these require an insertion of $G^{ia}$ or a derivative, which are not present in Eqs.~\eqref{eq:spin-flavor-h0} and \eqref{eq:spin-flavor-ha}.
This observation is in agreement with the relative ordering of the $\pi N \Delta$ couplings discussed in Ref.~\cite{gandorParityTimereversalViolating2024}.

First, the analogue of Eq.~\eqref{eq:g1} that couples the neutral pion to the isoscalar combination of the $\Delta$ is given by\footnote{We include a factor of $1/2 F_0$ in the matching relations such that the $g^{(\pi \Delta \Delta)}$ couplings have the same form as the $g^{(\pi \NN)}$ couplings.}
    \begin{align}
        \frac{\bar g^{(\pi \Delta \Delta)}_1}{2 F_0} & = \frac{4 B_0 \bar m \varepsilon}{F_0} \Nc \left( h_0^{(0)} + \frac{15}{4 \Nc^2} h^{(0)}_2 \right) \tilde \theta \, . \label{eq:g1_Delta}
    \end{align}
Thus, the difference between $\bar g^{(\pi \Delta \Delta)}_1$ and $\bar g_1^{(\pi \NN)}$ is $O(1/\Nc^2)$ and arises from the $J^2$ operator. 
For the couplings analogous to $\bar g_0^{(\pi \NN)}$, we have
    \begin{align}
        \frac{\bar g^{(\pi \Delta \Delta)}_0}{2 F_0} & = - \frac{6 a \tilde \theta}{F_0 \Nc} \left( h_1^{(v)} + \frac{15}{4 \Nc^2} h^{(v)}_3 \right) \, ,
    \end{align}
In the \sufour limit, these relations imply 
    \begin{align}
        \frac{g_0^{(\pi \Delta \Delta)}}{g_0^{(\pi \NN)}} = -3 + O(1/\Nc^2) \, , \\
        \frac{g_1^{(\pi \Delta \Delta)}}{g_1^{(\pi \NN)}} = 1 + O(1/\Nc^2) \, .
    \end{align}
These relations could provide useful benchmarks for future lattice calculations similar to the validation of large-\Nc mass relations on the lattice in Ref.~\cite{jenkinsLatticeTest12010}.
Additionally, they give a first quantitative estimate of the $\cancel{\rm CP}$ $\pi \Delta \Delta$ couplings that should enter the parity violating and time-reversal-invariance violating \NN potential, albeit at higher orders than recently considered \cite{gandorParityTimereversalViolating2024}.

\section{Electric dipole moments}
    \label{sec:edm}

The \gls{nEDM} has been calculated in several places in \gls{ChiPT} \cite{hockingsElectricDipoleForm2005, mereghettiElectricDipoleForm2011, guoBaryonElectricDipole2012, ottnadNewInsightsNeutron2010, pichStrongCPviolationEffective1991, borasoyElectricDipoleMoment2000}.
Here, we provide an independent check of the leading one-loop contribution to the nucleon EDMs.
Furthermore, we extend the analysis to the $\Delta$ as well as the $\Delta N$  transition moment via \sufour spin-flavor symmetry.
However, we will see that the large-\Nc analysis could provide a different perspective on the relative importance of tree-level and one-loop terms.

The loop contribution to the amplitude for the $B \gamma \to B'$ transition is given by
    \begin{align}
        i \calM^{(\rm loop)}_{B' B} & = e q^i \frac{4 a \tilde \theta}{F_0^2 \Nc} \epsilon^{3ab} \sum_{B_I} \frac{d}{d m_\pi^2} \left( I_\pi + \Delta_{B_I} J_{\pi B_I} \right) \nonumber \\
        & \times \left[ \calH^a_{B' B_I} \calP_{B_I} A^{ib}_{B_I B} - A^{ib}_{B' B_I} \calP_{B_I} \calH^a_{B_I B} \right] \, . \label{eq:loop_edm_general}
    \end{align}
The spin projectors are a linear combination of $\1$ and $J^2$ (their explicit form can be found in the Appendix) while $\calH^a \sim I^a$; therefore, these operators commute.
The first term in square brackets will then force the intermediate baryon to have the same spin as the outgoing baryon $B'$ while the second term will force the intermediate baryon to have the same spin as the incoming baryon $B$.
If the incoming and outgoing baryons are the same species, then the intermediate state can only have the same spin as the external baryons.
Specifically, if we consider $B = B' = N$, then there is no intermediate $\Delta$ contribution and vice-versa.
This can be understood from the fact that the only operator that can change the baryon spin at this order is the axial current $A^{ia}$.
In order to have a contribution from an intermediate $\Delta$ to the nucleon EDM, we would need two insertions of $A^{ia}$.
Therefore, the spin-diagonal matrix elements will not receive contributions from other baryons in the spin-tower until at least two-loop order or from higher-order operators in the chiral expansion.

In the case that both the incoming and outgoing baryons are the same spin, the EDM only receives a contribution from the term proportional to $I_\pi$ in Eq.~\eqref{eq:loop_edm_general}.
This is also true in the degeneracy limit $\Delta_{B_I} \to 0$.
Then the term in square brackets reduces to a commutator,
    \begin{align}
        i \calM^{(\rm loop)}_{B B} & = i e q^i \frac{4 a \tilde \theta}{\Nc} \left( \frac{1}{4 \pi F_0} \right)^2 \left[\frac{1}{\epsilon} + \log \left( \frac{\mu^2}{m_\pi^2} \right) \right] \nonumber \\
        & \times \epsilon^{3ab} \left[ \calH^a, A^{ib} \right]_{B B} \, .
    \end{align}
For our purposes, we only need the leading terms in the expansions of $\calH^a$ and $A^{ib}$ from Eqs.~\eqref{eq:spin-flavor-ha} and \eqref{eq:spin-flavor-axial}, respectively.
The \sufour algebra then reduces the commutator to a single operator,
    \begin{align}
        i \calM^{(\rm loop)}_{B B} & = -2 e q^i \frac{4 a \tilde \theta}{\Nc} \left( \frac{1}{4 \pi F_0} \right)^2 \left[\frac{1}{\epsilon} + \log \left( \frac{\mu^2}{m_\pi^2} \right) \right]  \nonumber \\
        & \times a_1^{(v)} h_1^{(v)} G^{i3}_{B B} \, , \label{eq:neutron-edm-degenerate1}
    \end{align}
which starts at $O(1/\Nc)$ and is valid up to relative $O(1/\Nc^2)$ corrections.
The tree-level contribution comes from an operator in the Lagrangian of the form
    \begin{equation}
            \label{eq:edm_tree}
        \calL \supset  \calD_v^{i3} E^i + \calD_s^i E^i \, , 
    \end{equation}
where $\calD_v^{i3}$ is an isovector electric dipole moment and $\calD_s^{i}$ is an isoscalar.
The baryon matrix elements of $\calD_v^{i3}$ have the same spin-flavor expansion as the isovector axial current in Eq.~\eqref{eq:spin-flavor-axial}, 
    \begin{align}
        \calD_v^{i3} & = d_1^{(v)} G^{i3} + \frac{1}{\Nc} d^{(v)}_2 J^i I^3 + \frac{1}{\Nc^2} d^{(v)}_3 \{ J^2, G^{i3} \} \, . 
    \end{align}
On the other hand, the isoscalar operator has the expansion
    \begin{align}
        \calD^i_s & = d_1^{(0)} J^i + O(1/\Nc^2) \, .
    \end{align}
All of the coefficients are $O(1)$ apart from some number of dimension ${\rm MeV}^{-1}$, and $d_1^{(v)}$ is renormalized to cancel the $1/\epsilon$ pole in Eq.~\eqref{eq:neutron-edm-degenerate1}.
Therefore, the leading order (with respect to the chiral expansion) renormalized expression through $O(1/\Nc)$ is
    \begin{align}
        d_B & = \left[ d_1^{(v)} + \left( \frac{1}{4 \pi F_0} \right)^2 \frac{8 a \tilde \theta}{\Nc} a_1^{(v)} h_1^{(v)} \log \frac{\mu^2}{m_\pi^2} \right] G^{i3}_{B B} \nonumber \\
        & + d_1^{(0)} J^i_{BB} + \frac{1}{\Nc} d_2^{(v)} (J^i I^3)_{BB} \, . \label{eq:edm_master}
    \end{align}
Here, $d_1^{(v)}$ runs such that the term in square brackets is $\mu$ independent.
It is clear in this form that the $d_1^{(v)}$ contribution is $O(\Nc)$ and the $d_1^{(0)}$ contribution is $O(\Nc^0)$ while the one-loop and $d_2^{(v)}$ contributions are both $O(1/\Nc)$; we have dropped the $O(1/\Nc^2)$ contributions.

Before proceeding to the individual moments, several comments regarding the relation of this expression to past work are in order.
First, consider the relative sizes of the tree-level and one-loop contributions in Eq.~\eqref{eq:edm_master}.
The first term in square brackets is na\"ively $O(1)$ while the second term is $O(1/\Nc^2)$.
This is a reflection of the fact that the spin-flavor symmetry turns the loop expansion of \gls{ChiPT} into a 1/\Nc expansion \cite{flores-mendietaStructureLargeCancellations2000}.
This relative scaling of tree-level and one-loop contributions differs from the conventional wisdom that
``the short- and long-range contributions are in general of the same size" \cite{hockingsElectricDipoleForm2005}.
A relative suppression of the loop has also been observed in Ref.~\cite{pospelovThetaVacuaQCD2000}, although they only find a $1/\Nc$ suppression.
This difference can be traced back to the singlet mass.
In Ref.~\cite{pospelovThetaVacuaQCD2000}, it enters in the form $1 - m_\pi^2/m_\eta^2$, which is $O(1/\Nc)$.
Here it enters through the factor $a/\Nc$, which leads to the extra $1/\Nc$ suppression when combined with the scaling $F_0 \sim \sqrt{\Nc}$ (although it is numerically significant since $a/\Nc \sim (600 \ {\rm MeV})^2$).

Second, Refs.~\cite{borasoyElectricDipoleMoment2000, ottnadNewInsightsNeutron2010} use large-\Nc arguments to estimate the relative sizes of couplings in the ${\rm U}(3)$ chiral Lagrangian.
However, neither of these estimates are based on the spin-flavor symmetry known to emerge in the large-\Nc limit.
While we agree concerning the numerical estimate of the loop contribution, they suggest that the tree-level contribution is at most $O(\Nc^0)$.
However, we see here that the tree-level contribution is at most $O(\Nc)$ because it is proportional to $G$.

Now, we consider the EDMs of the nucleons and the $\Delta$s.
The matrix elements of the spin-flavor generators are taken between $J_z = 1/2$ states for the nucleons and $J_z = 3/2$ states for the $\Delta$s.
For the neutron, we obtain
    \begin{align}
        d_n & = -\frac{5}{12} \left[ d_1^{(v)} + \left( \frac{1}{4 \pi F_0} \right)^2 \frac{8 a \tilde \theta}{\Nc} a_1^{(v)} h_1^{(v)}  \log \frac{\mu^2}{m_\pi^2}  \right]  \nonumber \\
        & + \frac{1}{2} d_1^{(s)} - \frac{1}{12 } d_2^{(v)} \, , \label{eq:neutron-edm-degenerate2}
    \end{align}
where we have evaluated this with $\Nc = 3$.
The loop contribution agrees with Ref.~\cite{hockingsElectricDipoleForm2005} when we use the leading order approximations $g_A = \frac{5}{6} a_1^{(v)}$ and the first term of Eq.~\eqref{eq:g0}, which we reiterate are valid up to relative $O(1/\Nc^2)$ corrections.
To obtain a numerical estimate of the loop contribution, we take $g_A = 1.2754(13)$ \cite{navasReviewParticlePhysics2024} and the nucleon mass splitting from Eq.~\eqref{eq:mass_params},
    \begin{equation}
        d_n^{(\rm loop)} = -0.0018(5) \ \bar \theta \ e \ \text{fm} \, , \label{eq:neutron_loop_estimate}
    \end{equation}
where the error is obtained by varying $\mu$ between 500 MeV and 1 GeV.
We have converted to units of $\bar \theta$ through the relation Eq.~\eqref{eq:aOverNc_pion_mass1}, and our estimate is consistent with Refs.~\cite{borasoyElectricDipoleMoment2000, ottnadNewInsightsNeutron2010, guoBaryonElectricDipole2012, pichStrongCPviolationEffective1991}. 
Since the loop contribution is relatively $1/\Nc^2$ suppressed, we estimate the tree-level contribution by multiplying Eq.~\eqref{eq:neutron_loop_estimate}by 10 and adding a $40\%$ uncertainty to account for all of the subleading terms in $1/\Nc$ expansion,
    \begin{equation}
        d_n \lesssim - 0.018(7) \, \bar \theta \, e \, {\rm fm} \, . \label{eq:nedm_bound}
    \end{equation}
However, this should only be interpreted as an upper bound.
Using the experimental bound $d_n^{(\rm expt)} \leq 1.8 \times 10^{-13} \, e \, {\rm fm}$ \cite{abelMeasurementPermanentElectric2020} results in the limit
    \begin{equation}
        \abs{\bar \theta} \lesssim 10^{-11} \, .
    \end{equation}
If we use the loop contribution instead, then we find $\abs{\bar \theta} \lesssim 10^{-10}$ in agreement with other determinations.

The estimate in Eq.~\eqref{eq:nedm_bound} is about one order of magnitude larger than the lattice calculations from Refs.~\cite{dragosConfirmingExistenceStrong2021, alexandrouNeutronElectricDipole2021, liangNucleonElectricDipole2023}, but it is compatible with the result of Ref.~\cite{bhattacharyaContributionQCD$Th$term2021}.
It is possible that this discrepancy can partially be attributed to the way chiral extrapolation is performed.
Specifically, we have already seen that it is possible for the one-loop contribution to the \gls{nEDM} to be $1/\Nc^2$ suppressed relative to the tree-level contribution.
The fit in Ref.~\cite{dragosConfirmingExistenceStrong2021}, for example, does not support this conclusion, nor does it fully rule out the possibility that there is some loop suppression within the quoted errors.
On the other hand, the chiral extrapolations from Ref.~\cite{bhattacharyaContributionQCD$Th$term2021} both with and without $N \pi$ excited states indicate that the loop contribution is about half the size of the tree-level contribution excluding the discretization effects.

We can also obtain analogous moments for the proton and the $\Delta$s by replacing the matrix elements of the spin-flavor generators in Eq.~\eqref{eq:neutron-edm-degenerate1},
    \begin{align}
        d_{p} & = \frac{5}{12} \tilde d_1^{(v)} + \frac{1}{2} d_1^{(s)} + \frac{1}{12} d_2^{(v)} \, , \\
        d_{\Delta^{++}} & = \frac{3}{4} \tilde d_1^{(v)} + \frac{3}{2} d_1^{(s)} + \frac{3}{4} d_2^{(v)} \, , \\
        d_{\Delta^+} & = \frac{1}{4} \tilde d_1^{(v)} + \frac{3}{2} d_1^{(s)} + \frac{1}{4} d_2^{(v)} \, , \\
        d_{\Delta^0} & = - \frac{1}{4} \tilde d_1^{(v)} + \frac{3}{2} d_1^{(s)} - \frac{1}{4} d_2^{(v)} \, , \\
        d_{\Delta^-} & = - \frac{3}{4} \tilde d_1^{(v)} + \frac{3}{2} d_1^{(s)} - \frac{3}{4} d_2^{(v)} \, ,
    \end{align}
where we have introduced the $\mu$ independent combination
    \begin{equation}
        \tilde d_1^{(v)} = d_1^{(v)} + \left( \frac{1}{4 \pi F_0} \right)^2 \frac{8 a \tilde \theta}{\Nc} a_1^{(v)} h_1^{(v)}  \log \frac{\mu^2}{m_\pi^2} \, .
    \end{equation}
Because the electric dipole operator has an expansion similar to that of the magnetic moment \cite{jenkinsBaryonMagneticMoments1994a}, we can obtain isospin combinations that could be tested with lattice data.
First, we would expect
    \begin{equation}
        \frac{d_p + d_n}{d_p - d_n} = 0 + O(1/\Nc) \, ,
    \end{equation}
which is corroborated by the existing lattice calculations.
A few additional relations that would provide a nontrivial cross-check of the $1/\Nc$ expansion are
    \begin{align}
        d_{\Delta^{++}} & = \frac{9}{10} \left( d_p - d_n \right) + \frac{3}{2} \left( d_p + d_n \right) \label{eq:Delta++_p-n} \, , \\
        d_{\Delta^{++}} - d_{\Delta^-} & = \frac{9}{5} \left( d_p - d_n \right) \label{eq:Delta++_Delta-}\, .
    \end{align}
The first relation is valid up to $O(1/\Nc^2)$ corrections while the second relation is likely more precise.
Moreover, these relations do not necessarily rely on the chiral expansion, thus they could be tested at heavier pion masses.

Finally, we consider an off-diagonal transition matrix element for $\Delta \gamma \to N$.
Here, one must make use of the operator reduction rules of Ref.~\cite{dashenSpinFlavorStructure1995}.
There is also a contribution from the $J_{\pi B}$ integral,
    \begin{align}
        & i \calM = - 2  q^i a_1^{(v)} h_1^{(v)} G^{i3}_{N \Delta} \frac{4 a \tilde \theta}{\Nc} \frac{1}{(4 \pi F_0)^2} 
        \left[ \frac{1}{\epsilon} + \log \left( \frac{\mu^2}{m_\pi^2} \right) \right. \nonumber \\
        & \left. - \frac{2 \Delta^2}{m_\pi^2} -  \frac{\Delta}{\sqrt{\Delta^2 - m_\pi^2}} \log \left( \frac{2(\Delta^2 - \Delta \sqrt{\Delta^2 - m_\pi^2}) - m_\pi^2}{m_\pi^2} \right) \right] \, , \label{eq:transitionM}
    \end{align}
where $\Delta = M_\Delta - M_N$.
Interestingly, this matrix element does not receive contributions from the higher-order operators in the spin-flavor expansion at this order in \gls{ChiPT}.

If we consider explicitly the transition $\Delta^+ \gamma \to p$, then the matrix elements of the spin-flavor generator will reduce this to
    \begin{align}
        & d_{\Delta^+ p} = \frac{\sqrt{2}}{3} a_1^{(v)} h_1^{(v)} \frac{8 a \tilde \theta}{\Nc} \frac{1}{(4 \pi F_0)^2} 
        \left[ d_1^{\rm EDM} + \log \left( \frac{\mu^2}{m_\pi^2} \right) \right. \nonumber \\
        & \left. - \frac{2 \Delta^2}{m_\pi^2}  -  \frac{\Delta}{\sqrt{\Delta^2 - m_\pi^2}} \log \left( \frac{2(\Delta^2 + \Delta \sqrt{\Delta^2 - m_\pi^2}) - m_\pi^2}{m_\pi^2} \right) \right] . \label{eq:Delta-proton-gamma}
    \end{align}
In the degeneracy limit $\Delta \to 0$, both terms on the second line vanish such that the result is directly proportional that of the neutron,
    \begin{equation}
        \left. d_{\Delta^+ p} \right|_{\Delta = 0}   = - \frac{4 \sqrt{2}}{5} d_n \approx -1.13 \, d_n
    \end{equation}
For the physical case $\Delta = 293$ MeV, the loop contribution yields
    \begin{equation}
        \left. d^{(\rm loop)}_{\Delta^+ p} \right|_{\Delta = 293 \, {\rm MeV}}   = 0.9(4) d_n \, ,
    \end{equation}
where the error mostly comes from the variation of $\mu$ between 500 and 1000 MeV.
There are additional isospin relations that could be tested on the lattice,
    \begin{align}
        d_{\Delta^+ p} & = d_{\Delta^0 n} \, , \label{eq:transition1} \\
        d_{\Delta^+ p} + d_{\Delta^0 n} & = \frac{4 \sqrt{2}}{5} (d_p - d_n) + O(1/\Nc^2) \label{eq:transition2}  \, .
    \end{align}
Again, these relationships should hold even at larger pion masses when the nucleon and $\Delta$ are closer to being degenerate.

\section{Conclusion}
    \label{sec:conclusion}

In this work, we have used the spin-flavor symmetry of the baryon sector from the 1/\Nc expansion to reassess the $\cancel{\rm CP}$ pion-nucleon couplings and EDMs.
We have also derived new constraints for the $\cancel{\rm CP}$ pion-$\Delta$ couplings and EDMs as well as a $\Delta N$ transition moment.

The constraints for the pion-nucleon couplings largely agree with previous work.
However, the pion-$\Delta$ couplings are so far unconstrained to the best of our knowledge.
This work fills in that gap by relating these couplings to the pion-nucleon couplings.
To obtain numerical estimates, we have also employed the Witten-Veneziano fomula with lattice input for the topological susceptibility.
This extends the idea set forth in Ref.~\cite{devriesLatticeQCDSpectroscopy2017} to make use of baryon sepctroscopy on the lattice to constrain $\cancel{\rm CP}$ properties of \gls{ChiPT}; specifically, we can make use of lattice results from pure Yang-Mills to constrain the isoscalar pion-baryon couplings.
In principle, these constraints could be implemented in future generations of parity- and time-reversal-invariance-violating \NN potentials.

The utility of the spin-flavor expansion is highlighted by Eq.~\eqref{eq:edm_master}, which is a universal result in the sense that it applies to all $\cancel{\rm CP}$ nucleon and $\Delta$ matrix elements.
Again, our results for the loop contributions to the nucleon EDMs are in good agreement with other analyses.
One difference is the observation that the loop contribution is formally $1/\Nc^2$ suppressed relative to the tree-level contribution, although the tree-level results should be considered as an upper bound rather than a strict prediction.
Current lattice calculations \cite{dragosConfirmingExistenceStrong2021, liangNucleonElectricDipole2023, alexandrouNeutronElectricDipole2021} neither support this enhancement nor fully rule it out.
On the other hand, our results appear to be consistent with the results of Ref.~\cite{bhattacharyaContributionQCD$Th$term2021}.
Regardless of this discrepancy, the large-\Nc constraints for the relative size of the isoscalar and isovector combinations of the neutron and proton EDMs appear to be in good agreement with the available lattice data.

We have also provided novel results for $\Delta$ EDMs and $\Delta$-nucleon transition moments.
There are also various combinations of these such as Eqs.~\eqref{eq:Delta++_p-n}, \eqref{eq:Delta++_Delta-}, \eqref{eq:transition1}, and \eqref{eq:transition2} that could be tested on the lattice even at heavy pion masses where the $\Delta$ is a stable particle.
If these matrix elements can be accessed phenomenologically, then they would provide a new avenue for constraining $\bar \theta$.

\begin{acknowledgments}
    I would like to thank Kaori Fuyuto, Emanuele Mereghetti, and Andr\'e Walker-Loud for stimulating discussions.
    This work was supported by the NSF through cooperative agreement 2020275 and by the DOE Topical Collaboration “Nuclear Theory for New Physics,” award No. DE-SC0023663.
\end{acknowledgments}


\onecolumngrid
\appendix

\section{Loop integrals}

The loop integrals needed for the loop contribution to the EDMs are
    \begin{align}
        \begin{split}
            I_\pi(m^2) & = \mu^{4-n} \int \frac{d^n k}{(2 \pi)^n} \frac{1}{k^2 - m^2 + i \epsilon} \, , \\
            & = i \left( \frac{m}{4 \pi} \right)^2 \left[ \frac{1}{\epsilon} + \log \left( \frac{\tilde \mu^2}{m^2} \right) + 1 \right] \, , 
        \end{split} \\
        J_{\pi B}(m^2, \omega) & =  \mu^{4-n} \int \frac{d^n k}{(2 \pi)^n} \frac{1}{k^2 - m^2 + i \epsilon} \frac{1}{v \cdot k + \omega + i \epsilon} \, , \nonumber \\
        & = 
        \begin{cases}
            \frac{2i}{(4 \pi)^2} \left[ \omega \left( \frac{1}{\epsilon} + 2 + \log \left( \frac{\tilde \mu^2}{m^2} \right) \right) - 2 \sqrt{m^2 - \omega^2} \left( \frac{\pi}{2} - \tan^{-1} \left( \frac{\omega}{\sqrt{m^2 - \omega^2}} \right) \right) \right] \, ,& m^2 > \omega^2 \, , \\
            \frac{2i}{(4 \pi)^2} \left\{ \omega \left[ \frac{1}{\epsilon} + 2 + \log \left( \frac{\tilde \mu^2}{m^2} \right) \right] - \sqrt{\omega^2 - m^2} \log \left( \frac{\omega + \sqrt{\omega^2 - m^2}}{\omega - \sqrt{\omega^2 - m^2}} \right) \right\} \, , & \omega^2 > m^2 \, .
        \end{cases}
    \end{align}

\section{Spin-flavor matrix elements}

The matrix elements of the spin-flavor generators needed in this work are,
    \begin{align}
        \bra{B'} J^i \ket{B} & = \sqrt{2 S + 1} \braket{S', S_z'; I', I_z'}{S, S_z; 1, i} \delta_{I' I} \delta_{I_3' I_3} \, , \\
        \bra{B'} I^a \ket{B} & = \sqrt{2 I + 1} \braket{I', I_z'}{I, I_z; 1, a} \delta_{S' S} \delta_{S_z' S_z} \, , \\
        \bra{B'} G^{ia} \ket{B} & = \frac{1}{4} \sqrt{\frac{2 S + 1}{2 S' + 1}} \sqrt{(2 + \Nc)^2 - (S - S')^2 (S + S' + 1)^2} \braket{S', S_z'; I', I_z'}{S, S_z; 1, i} \braket{I', I_z'}{I, I_z; 1, a} \, ,
    \end{align}
where $S$ and $I$ ($S'$ and $I'$) are the spin and isospin of baryon $B$ ($B'$) with projections $S_z$ and $I_z$ ($S_z'$ and $I_z$'), respectively.

The spin projection operators for intermediate baryons are
    \begin{align}
        \calP_{1/2} & = \frac{5}{4} - \frac{1}{3} J^2 \, , \\
        \calP_{3/2} & = - \frac{1}{4} + \frac{1}{3} J^2 \, ,
    \end{align}
such that baryon propagators with spin $j$ are written as
    \begin{equation}
        \frac{i \calP_j}{k_0 - \Delta_j}
    \end{equation}
with 
    \begin{align}
        \Delta_{\frac{1}{2}} & = \begin{cases}
            0,  & j_{\rm in} = 1/2  \\
            - \Delta, & j_{\rm in} = 3/2 
        \end{cases} \\
        \Delta_{\frac{3}{2}} & = \begin{cases}
            \Delta,  & j_{\rm in} = 1/2  \\
            0, & j_{\rm in} = 3/2 
        \end{cases} \, ,
    \end{align}
for an incoming baryon with spin $j_{\rm in}$. The nucleon-$\Delta$ mass splitting is denoted by $\Delta = M_\Delta - M_N$.

\bibliography{CP_violation_refs}
\end{document}